\documentclass{article}
\usepackage{spconf,amsmath,graphicx}

\usepackage{cite}
\usepackage{enumitem}
\usepackage{amssymb}
\usepackage{multirow}
\usepackage{amsfonts}
\usepackage{bbding}
\usepackage[fontsize=9pt]{fontsize}

\title{Push-Pull: Characterizing the Adversarial Robustness for Audio-Visual \\ Active Speaker Detection}

\name{BLIND}
\address{BLIND}
\email{}

\name{Xuanjun Chen$^{1*}$, Haibin Wu$^{23*}$, Helen Meng$^{34\dagger}$, Hung-yi Lee$^{2\dagger}$, Jyh-Shing Roger Jang$^{1\dagger}$\thanks{${*}$ Equal contribution. ${\dagger}$ Equal correspondence.}}

\address{
  $^1$Department. of Computer Science Information Engineering, National Taiwan University \\
  $^2$Graduate Institute of Communication Engineering, National Taiwan University \\
  $^3$Centre for Perceptual and Interactive Intelligence, The Chinese University of Hong Kong \\
  $^4$Human-Computer Communications Laboratory, The Chinese University of Hong Kong}
\email{\{r09922165, f07921092, hungyilee\}@ntu.edu.tw, hmmeng@se.cuhk.edu.hk,jang@csie.ntu.edu.tw}

\begin{document}

\maketitle

\begin{abstract}

Audio-visual active speaker detection (AVASD) is well-developed, and now is an indispensable front-end for several multi-modal applications. 
However, to the best of our knowledge, the adversarial robustness of AVASD models hasn't been investigated, not to mention the effective defense against such attacks.
In this paper, we are the first to reveal the vulnerability of AVASD models under audio-only, visual-only, and audio-visual adversarial attacks through extensive experiments. 
What's more, we also propose a novel audio-visual interaction loss (AVIL) for making attackers difficult to find feasible adversarial examples under an allocated attack budget. 
The loss aims at pushing the inter-class embeddings to be dispersed, namely non-speech and speech clusters, sufficiently disentangled, and pulling the intra-class embeddings as close as possible to keep them compact. 
Experimental results show the AVIL outperforms the adversarial training by 33.14 mAP (\%) under multi-modal attacks.

\end{abstract}

\keywords{Audio-visual active speaker detection, multi-modal adversarial attack, adversarial robustness}

\section{Introduction}
Active Speaker Detection (ASD) seeks to detect who is speaking in a visual scene containing one or more speakers \cite{49517, roth2020ava}.
Recently, audio-visual ASD (AVASD), which integrates audio-visual information by learning the relationship between speech and facial motion, effectively improves the performance of ASD, and AVASD has become more indispensable as a front-end for multi-modal applications.
However, to the best of our knowledge, whether the AVASD models are robust against adversarial attacks has not been investigated previously, not to mention the effective defense method against such multi-modal attacks.

Crafting indistinguishable adversarial noise, adding such noise to clean samples to generate adversarial samples, and then manipulating the AI models by such samples, is called \emph{adversarial attack} \cite{szegedy2013intriguing}.
Previous adversarial attacks usually focus on single-modal applications.
For visual-modal attacks, Szegedy et al. first propose to attack state-of-the-art image classification models \cite{szegedy2013intriguing} in 2013.
For the speech modality, models including automatic speaker verification (ASV) systems \cite{kreuk2018fooling,wu2021adversarial,jati2021adversarial,wu2021voting,xie2021real,xie2020real,abdullah2021sok,marras2019adversarial,wu2021improving,das2020attacker,wu2022adversarial,li2020practical,peng2021pairing,zhang2021attack,tan2022adversarial}, anti-spoofing models for ASV \cite{liu2019adversarial,wu2020defense2,zhang2020adversarial,kassis2021practical,wu2020defense}, and automatic speech recognition models \cite{carlini2018audio,yakura2018robust,taori2019targeted,qin2019imperceptible,alzantot2018did,schonherr2018adversarial,yang2020characterizing} are also vulnerable to adversarial attacks.
For audio-visual learning, Li et al. \cite{li2022adversarial} studied the audio-visual adversarial robustness of the general sound event detection model but only considered single- or multi-modal attacks under an attack method.

Given that AVASD is now ubiquitously implemented as a front-end for a variety of multi-modal downstream models, the dangerous adversarial noise may manipulate the AVASD front-end to commit errors, which will accumulate and propagate to the downstream applications. 
Hence it is of high priority that we mitigate the adversarial vulnerability of AVASD and ensure robustness against such attacks.
This paper investigates the susceptibility of AVASD models to adversarial attacks and then proposes a novel defense method to improve their robustness.
Our contributions are summarized in two folds: 1). To the best of our knowledge, this is the first work to reveal the vulnerability of AVASD models under three kinds of attacks, including audio-only, visual-only, and audio-visual adversarial attacks by extensive experiments. 2). We also propose a novel audio-visual interaction loss (AVIL), which aims at pushing the inter-class embeddings, namely the non-speech and speech clusters, sufficiently disentangled, and pulling the intra-class embeddings as close as possible. Expanding the inter-class dispersion and enhancing the intra-class compactness will make it difficult for attackers to find feasible adversarial samples to go beyond the decision boundary within the allocated attacking budget. The experimental results illustrate that the brand-new audio-visual interaction loss effectively strengthens the invulnerability of AVASD models.

\begin{figure*}
\centering
\includegraphics[width=17.5cm]{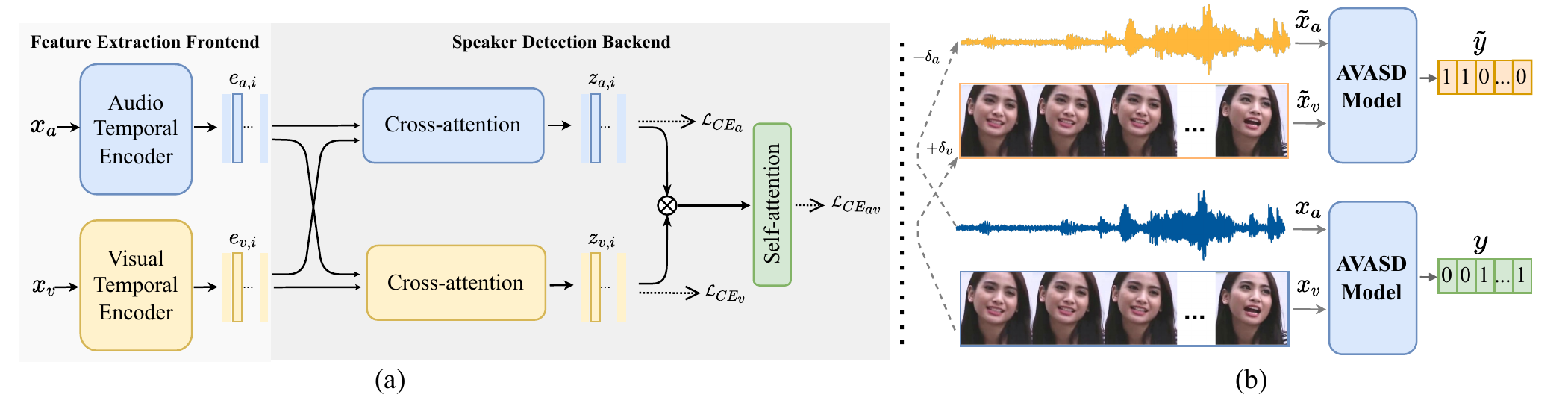}
\vspace{-1.3em}
\caption{(a) The TalkNet framework. $x_a$ and $x_v$ are the audio and visual inputs, respectively. $\otimes$ denotes the concatenation procedure. $\mathcal{L}_{CE_{a}}$, $\mathcal{L}_{CE_{v}}$ and $\mathcal{L}_{CE_{av}}$ are the cross entropy losses for audio-only prediction head, visual-only prediction head, and audio-visual prediction head, respectively. (b) The audio-visual attack framework for AVASD. $x_a$ and $x_v$ are the audio and visual samples respectively, $y$ is the ground-truth for the multi-sensory input $\{x_a, x_v\}$.  $\delta_a$ and $\delta_v$ are the adversarial perturbations for $x_a$ and $x_v$, respectively. $\tilde{y}$ is the prediction for the adversarial samples $\{\tilde{x}_a, \tilde{x}_v\}$. The adversarial attack aims at maximizing the difference between $y$ and $\tilde{y}$.}
\label{fig:system_overview}
\end{figure*}

\section{Background}

\subsection{Audio-Visual Active Speaker Detection}
The ASD task has been studied using audio, video, or the fusion of both. 
For audio, the voice activity detector \cite{ding2019personal, sehgal2018convolutional} is often used to detect the presence of speech. 
However, in real-world scenarios, the speech signal from the microphones is easily mixed with overlapping speech and background noise, which will hinder the effectiveness of voice activity detection. 
The visual part \cite{chakravarty2015s, shahid2021s} mainly analyzes the face and upper body of a person to determine whether the person is speaking, but the performance is limited due to some non-speech activities, e.g. licking lips, eating, and grinning.
The audio-visual processing refers to the combination of audio and visual parts \cite{ephrat2018looking, minotto2015multimodal}, and allows learning across modalities about the relationship between audio speech and facial motions. With valuable support from sizeable datasets, e.g. AVA-Active Speaker, and the AVA Challenge series  launched since 2018, a variety of high-performance models for AVASD have emerged recently \cite{kopuklu2021design, roy2021learning, zhang2021unicon,pouthier2021active,afouras2021sub,alcazar2021maas,chung2019naver,alcazar2020active,leon2021maas,zhang2019multi,tao2021someone}.
In real-world user authentication systems, AVASD can be used as a front-end task to assure security verification for speaker verification \cite{mak2020machine}. For AVASD system, there are four typical cases, such as speech without target speaker, no audible speaker, speaker without speech and speech with the target speaker. Only the speech with target speaker is labeled as speaking. Attackers possibly use some single modal attack methods or combine them to make AVASD produce wrong predictions in the other three cases, which is dangerous.
However, people have not yet seen investigations on the adversarial robustness of AVASD model. 

\subsection{Adversarial Attacks}
Adversarial attack is to manipulate a well-trained model to give wrong predictions by an adversarial sample, which is imperceptible by humans, compared with the original (unmodified) counterpart.
Mathematically, given a clean sample $x$ and the ground-truth label $y$, attack algorithms seek to find a sufficiently small perturbation $\delta$ such that: $\tilde{x} = x + \delta$, where $\tilde{x}$ is the adversarial sample that can fool the model to produce the wrong prediction $\tilde{y}$.
We can find a suitable $\delta$ by solving the following objective function:
\begin{equation}
\label{eq:attack_objective_equation}
\begin{aligned}
\mathop{\arg\max}_{\delta} \mathcal{L} (\tilde{x}, y, \theta), \\
s.t. || \delta ||_p \leq \epsilon,
\end{aligned}
\end{equation}
where $\mathcal{L}(\cdot)$ denotes the objective function pre-defined by the attackers, which is usually set to maximize the difference between $y$ and the model's final prediction given $\tilde{x}$, $\epsilon$ is the allowed perturbation budget, and $||\cdot||_p$ denotes the $p$-norm, which is usually considered to be a $l_{\infty}$-ball or $l_{2}$-ball centered at $x$. 
We evaluate the AVASD models' vulnerability with $l_{\infty}$-boundary adversarial noise, as it is widely used as a standard evaluation boundary for adversarial robustness \cite{madry2017towards}.
To solve the optimization problem as shown in Equation \ref{eq:attack_objective_equation}, we choose three widely used attack methods to evaluate the robustness of AVASD models, and the details are summarized below.

\noindent
\textbf{Basic Iterative Method (BIM)} BIM \cite{kurakin2016adversarial} is a method with iterative updates as follows:
\begin{equation}
\label{eq:Delta_equation}
\begin{aligned}
x_m = clip_\epsilon(x_{m-1} + \alpha \cdot \text{sign}( {\nabla_{x_{m-1}}} \mathcal{L}(x_{m-1}, y, \theta))), \\ for \  m = 1,..., M,
\end{aligned}
\end{equation}
where $x_0$ starts from the original sample $x$, $\alpha$ is the step size, $M$ is the number of iterations and the $clip_\epsilon(\cdot)$ function applies element-wise clipping to make $||x_{m-1} - x||_\infty \leq \epsilon, \epsilon \geq 0 \in \mathbb{R}$. 
The perturbed example $x_M$ is the final adversarial example.

\noindent
\textbf{Momentum-based Iterative Method (MIM)} MIM \cite{dong2018boosting} is an improved version of BIM. 
MIM introduces a momentum term into the iterative process to avoid BIM falling into local minimum and thus improve the attack performance over BIM.

\noindent
\textbf{Projected Gradient Descent (PGD)} PGD \cite{madry2017towards} is also a variant of BIM. 
PGD randomly initializes the adversarial noise $\delta$ for $\gamma$ times and conducts BIM-style attacks to generate $\gamma$ candidates of adversarial noise. 
Finally, the best one out of the $\gamma$ candidates with the best attack performance, will be chosen as the final adversarial sample.

\begin{figure*}
\centering
\includegraphics[width=17.5cm]{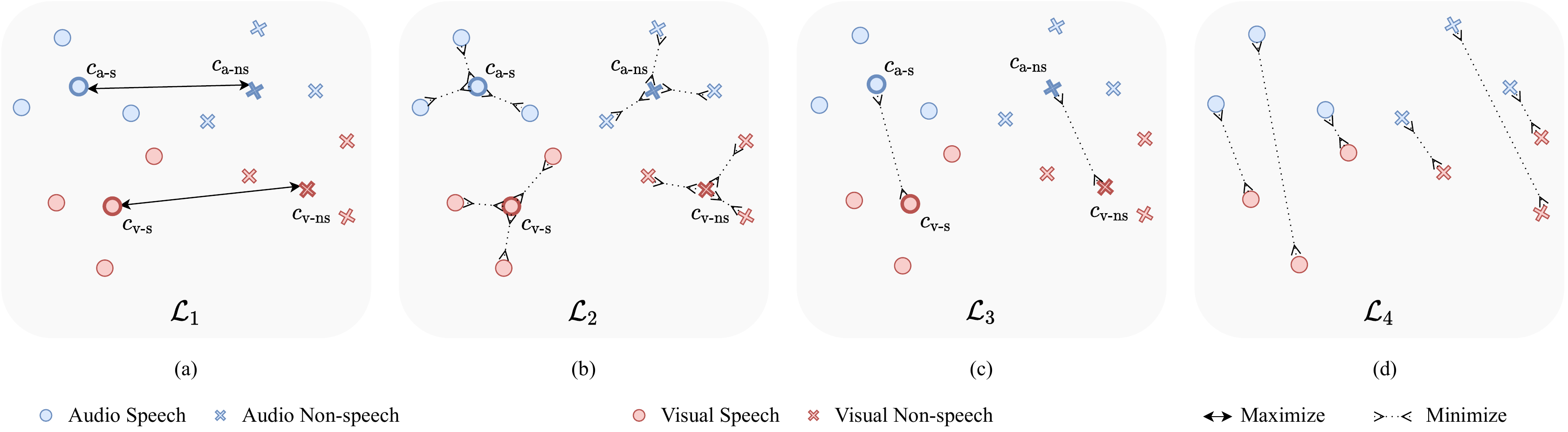}
\vspace{-1.3em}
\caption{The Audio-Visual Interaction Loss. The circle and cross fork denote the speech and non-speech embeddings, respectively. The colors blue and red present the audio and visual embeddings, respectively. The centers are those with bold borders.}
\label{fig:embeddingloss}
\end{figure*}

\section{Methodology}

\subsection{AVASD Model -- TalkNet}
We adopt TalkNet \cite{tao2021someone} for our case study to characterize the adversarial robustness of AVASD.
TalkNet is one of the state-of-the-art models for AVASD, which is fully end-to-end.
TalkNet takes a sequence of video frames $x_v$ consisting of cropped face sequences and the corresponding audio sequences $x_a$ as inputs.
The output probability denotes how likely the person is speaking in the given video frame.
TalkNet comprises a feature representation front-end, and a speaker detection back-end classifier, as shown in Fig.~\ref{fig:system_overview}.(a).
The front-end consists of an audio temporal encoder and a video temporal encoder to extract audio embeddings $e_{a,i}$ and visual embeddings $e_{v,i}$ for the $i^{th}$ frame.  
In the back-end, the audio and visual embeddings are aligned via inter-modality cross-attention and then concatenated to obtain the joint audio-visual embeddings $z_{a,i}$ and $z_{v,i}$ for the $i^{th}$ frame.
Then, a self-attention network is applied after the cross-attention network to model the audio-visual temporal information. 
Finally, a fully-connected layer with a softmax is implemented to project the output of the self-attention network to a sequence of ASD labels. 
The predicted label sequence is compared with the ground-truth label sequence by cross-entropy loss ($\mathcal{L}_{CE_{av}}$):
\begin{equation}
\label{eq:CE av}
\begin{aligned}
\mathcal{L}_{CE_{av}} = -\frac{1}{T}\sum_{t=1}^{T} (y_{t} \cdot log s_{t} + (1-y_{t}) \cdot log(1-s_{t})),
\end{aligned}
\end{equation}
where $y_t$ and $s_t$ are the ground-truth and predicted score for the $t^{th}$ frame, and $T$ is the total frames for one sample of video data. 
During training, TalkNet utilizes two additional predict heads for audio embeddings and visual embeddings after the cross-attention module to predict the ASD label sequences, as shown in the part of the speaker detection back-end in Fig. \ref{fig:system_overview}.(a).
The additional outputs here are used to calculate the weighted loss, and the final training loss is shown as follows:
\begin{equation}
\label{eq:CE all}
\begin{aligned}
\mathcal{L}_{CE_{all}} = \mathcal{L}_{CE_{av}} + 0.4 \times \mathcal{L}_{CE_{a}}  + 0.4 \times \mathcal{L}_{CE_{v}},
\end{aligned}
\end{equation}
where $\mathcal{L}_{CE_{a}}$ and $\mathcal{L}_{CE_{v}}$ denote the losses of audio-only and visual-only prediction head, respectively. The coefficient 0.4 is referred from the TalkNet \cite{tao2021someone}. During inference, only the prediction head after self-attention will be utilized. For further details of the above setting, please refer to the TalkNet paper \cite{tao2021someone}.

\subsection{Multi-Modal Attacks}

Let $x_a$ be an audio input, $x_v$ be the visual input and $y$ be the corresponding ground-truth label for the multisensory input: \{$x_a$, $x_v$\}.
We divide the audio-visual adversarial attack into three categories: 
the audio-only attack generates the audio adversarial example $\tilde{x}_a$,
the visual-only attack generates the visual adversarial example $\tilde{x}_v$ and
 the audio-visual attack generates multi-modal adversarial examples: {$\tilde{x}_a$, $\tilde{x}_v$}.
To force a well-trained audio-visual model to make wrong predictions with corresponding perturbations being as imperceptible as possible, the objective function for multi-modal attacks is as follows:
\begin{equation}
\label{eq:adversarial_objective}
\begin{aligned}
\mathop{\arg\max}_{\delta_a, \delta_v} \mathcal{L}(\tilde{x}_a, \tilde{x}_v, y), \\
s.t. \  || \delta_a ||_p \leq \epsilon_a, \ \ || \delta_v ||_p  \leq \epsilon_v,
\end{aligned}
\end{equation}
where $\tilde{x}_a = x_a + \delta_a$, $\tilde{x}_v = x_v + \delta_v$, $\mathcal{L}(\cdot)$ is the objective function to make the outputs of the audio-visual model as different as possible to $y$, $|| \cdot ||_p$ is the $p$-norm, and $\epsilon_a$ and $\epsilon_v$ are audio and visual perturbation budgets. 
In the case of an audio-only attack, the perturbation budget $\epsilon_v$ is equal to 0, and in the case of a visual-only attack, the perturbation budget $\epsilon_a$ is equal to 0. 
In the case of audio-visual attacks, both audio and visual inputs will be perturbed. Fig.~\ref{fig:system_overview}(b) illustrates the audio-visual adversarial attack framework.
Different strategies to search for $\delta$, which consists of $\delta_a$ and $\delta_v$, result in different adversarial attack methods.
Note that our multi-modal attack is jointly optimized on audio-visual modality instead of optimizing independently.
This paper adopts three famous attack methods for their effective attacking performance and affordable execution time based on our resources: BIM, MIM, and PGD.

We also set up two attack scenarios: training-aware attack and inference-aware attack scenarios.
In both kinds of attack scenarios, the attackers have full access to the model internals, including model architectures, parameters, and gradients.
Besides, the inference-aware attackers know exactly the inference procedure of the AVASD model.
In other words, they know the prediction head adopted for inference is the audio-visual head after the self-attention as shown in Fig.~\ref{fig:system_overview}(a), and then they will conduct adversarial attacks based on the loss as shown in Equation~\ref{eq:CE av}.
The inference-aware attack scenario is more practical, as it relies on the real inference procedure.
For the training-aware attackers, they even know the training loss of the AVASD model, and they will conduct adversarial attacks by Equation~\ref{eq:CE all}.
Training-aware attacks can craft even more dangerous attacks as they adopt all three prediction heads to find the adversarial perturbation.
Unless specified otherwise, all the experiments are conducted under the training-aware attack scenario as it is more dangerous.
We also perform the experiments for the inference-aware scenario and it shows the same trend. We show comparison results between training-aware and inference-aware attacks in Section~\ref{sec:infer vs train}.

\subsection{Audio-Visual Interaction Loss}
\label{sec:avil}

In this section, we first introduce the proposed audio-visual interaction loss (AVIL) and the implementation details.
Then we will present the rationale of the proposed method.

\noindent
\textbf{Implementation Procedure of AVIL.}
Suppose we have $K$ frames for one batch, and let $K_s$ and $K_n$ be the speech and non-speech frame numbers, respectively.
Let $\mathbb{S}$ and $\mathbb{N}$ denote the index sets for speech and non-speech.
We can get the four centers as below:
\begin{equation}
\label{eq:four centroid}
\begin{aligned}
&c_{a\text{-}s} = \frac{1}{K_s} \sum_{i \in \mathbb{S}} e_{a, i} \qquad 
&c_{a\text{-}ns} = \frac{1}{K_n} \sum_{i \in \mathbb{N}} e_{a, i} \\
&c_{v\text{-}s} = \frac{1}{K_s} \sum_{i \in \mathbb{S}} e_{v, i} \qquad
&c_{v\text{-}ns} = \frac{1}{K_n} \sum_{i \in \mathbb{N}} e_{v, i},
\end{aligned}
\end{equation}
where $c_{a\text{-}s}$, $c_{a\text{-}ns}$, $c_{v\text{-}s}$, $c_{v\text{-}ns}$ denote the centers for audio speech embeddings, audio non-speech embeddings, visual speech embeddings, visual non-speech embeddings, respectively. The centers are denoted with bold borders as shown in Fig.~\ref{fig:embeddingloss}. 


Then we can define the four audio-visual interaction losses:
\begin{itemize}
    \item Intra-modality inter-class dispersion (Fig.~\ref{fig:embeddingloss}.(a)): 
    \begin{equation}
    \label{loss1}
    \begin{aligned}
    \mathcal{L}_{1} = cos(c_{a\text{-}s}, c_{a\text{-}ns}) + cos(c_{v\text{-}s}, c_{v\text{-}ns}),
    \end{aligned}
    \end{equation}
    where $cos$ denotes the cosine similarity.
    
    \item Intra-modality intra-class dissimilarity (Fig.~\ref{fig:embeddingloss}.(b)):
    \begin{equation}
    \label{loss2}
    \begin{aligned}
    \mathcal{L}_{2} = & - (\frac{1}{K_s} \sum_{i \in \mathbb{S}} (cos(c_{a\text{-}s}, e_{a,i}) + cos(c_{v\text{-}s}, e_{v,i})) \\
                      & + \frac{1}{K_n} \sum_{i \in \mathbb{N}} (cos(c_{a\text{-}ns}, e_{a,i}) + cos(c_{v\text{-}ns}, e_{v,i})))
    \end{aligned}
    \end{equation}
    
    \item Inter-modality intra-class dissimilarity (center-based) as shown Fig.~\ref{fig:embeddingloss}.(c):
    \begin{equation}
    \label{loss3}
    \begin{aligned}
    \mathcal{L}_{3} = - ( cos(c_{a\text{-}s}, c_{v\text{-}s}) + cos(c_{a\text{-}ns}, c_{v\text{-}ns}) )
    \end{aligned}
    \end{equation}
    
    \item Inter-modality intra-class distance (sample-based) as shown in Fig.~\ref{fig:embeddingloss}.(d):
    \begin{equation}
    \label{loss4}
    \begin{aligned}
    \mathcal{L}_{4} = \frac{1}{K_s} \sum_{i \in \mathbb{S}} ||e_{v,i} - e_{a,i}||_2
                    + \frac{1}{K_n} \sum_{i \in \mathbb{N}} ||e_{v,i} - e_{a,i}||_2,
    \end{aligned}
    \end{equation}
    where $e_{a,i}$ and $e_{v,i}$ denote the speech embeddings for the audio and visual modalities, respectively. 
    When $1 \leq i \leq K $, $e_{a,i}$ and $e_{v,i}$ denote the non-speech embeddings.
\end{itemize}
To alleviate the adversarial noise, the four above losses mentioned above are adapted in the training process of the AVASD model and the final objective function is formulated as:
\begin{equation}
\label{eq:avrl objective}
\begin{aligned}
\mathcal{L}_{avil} = \mathcal{L}_{CE_{all}} + \sum_{j=1}^{4} \lambda_{j} \cdot \mathcal{L}_{j},
\end{aligned}
\end{equation}
where $\lambda_{j}$ denotes the hyperparameter.
Note that if $\{\lambda_j=0 \ for j=1,2,3,4\}$, $\mathcal{L}_{avil}$ will reduce to $\mathcal{L}_{CE_{all}}$, the training loss of the original TalkNet.
For training the model with AVIL, we simply set $\lambda_{1}=\lambda_{4}=\lambda_{2}=\lambda_{3}=0.1$, for simplicity.
And also, we just want to show the effectiveness of the four audio-visual interaction losses, rather than exhaust the hyperparameter settings and trickly improve the defense performance.


\noindent
\textbf{Rationale of AVIL.}
The adversarial attacks threaten the AVASD model by maximizing the loss functions, e.g. Equation~\ref{eq:CE av} and Equation~\ref{eq:CE all}, and then will urge the output far away from its original decision region \cite{szegedy2013intriguing, moosavi2016deepfool}.
For example, after the adversarial attack, the output for a speech frame will go away from the right region, namely ``speech'', and become non-speech.
As a result, it is reasonable that high inter-class dispersion and intra-class compactness will boost models' invulnerability, as it will make it hard for the attackers to find feasible adversarial perturbations within a given budget to push the genuine samples to pass through the decision boundary.

Minimizing $\mathcal{L}_{1}$ will equip the model with better discrimination capacity between speech and non-speech embeddings, resulting in higher inter-class difference from the models' perspective.
Maximizing $\mathcal{L}_{2}, \mathcal{L}_{3}$ and minimizing $\mathcal{L}_{4}$ will force the model to render more compact intra-class features.
Incorporating $\mathcal{L}_{1}, \mathcal{L}_{2}, \mathcal{L}_{3}, \mathcal{L}_{4}$ in the training process, we can simultaneously urge the model to learn both discriminative inter-class features, and compact intra-class features, leading the model less susceptible to adversarial perturbations.
As shown in Table~\ref{tab:ablation_study}, the four losses achieve the goal of significantly improving the robustness of the models.


\begin{figure*}
\centering
\includegraphics[width=16cm]{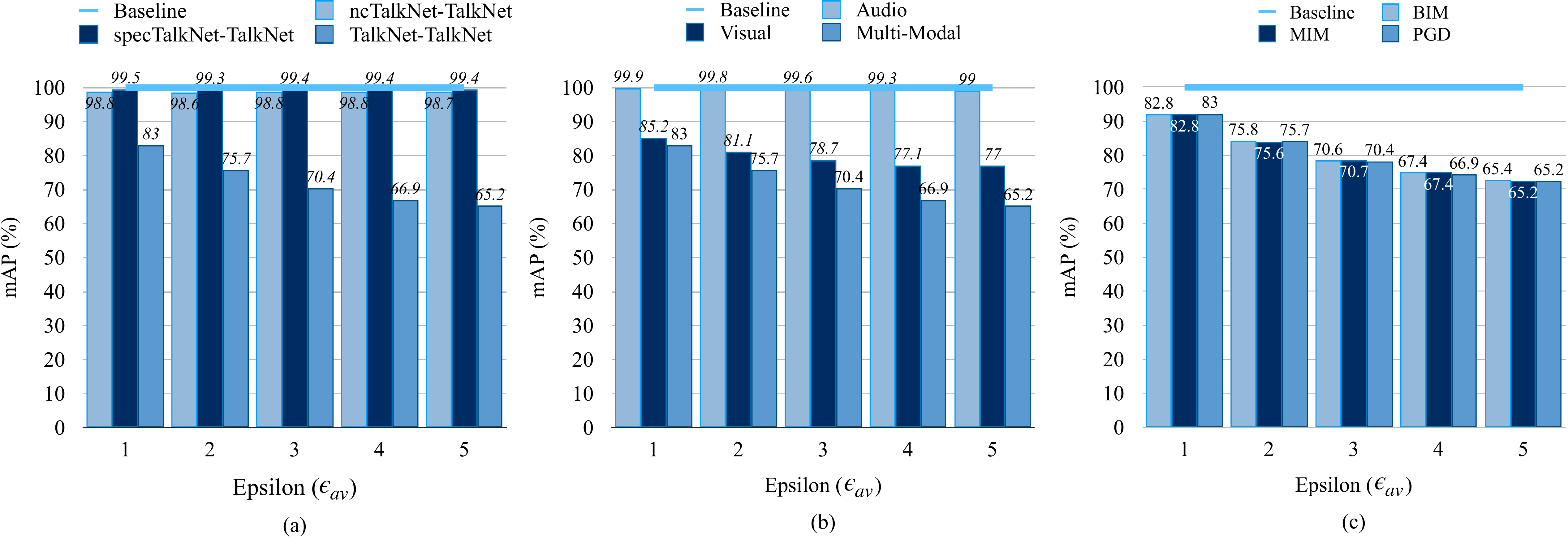}
\vspace{-1.3em}
\caption{Adversarial attack performance of AVASD models. (a) White-box and black-box attackers under multi-modal attack with PGD method. (b) Single-modal and multi-modal attack under white-box attacker with PGD method. (c) Different attack algorithms under white-box attacker with multi-modal attack. The attack budgets of audio and visual modals are $\epsilon_a = \epsilon_{av} \times 10^{-4}$ and $\epsilon_v = \epsilon_{av} \times 10^{-1}$, respectively.}
\label{fig:multi-attack-ablation}
\vspace{-0.5em}
\end{figure*}

\section{Experiment}
\subsection{Experimental setup}

We use TalkNet \cite{tao2021someone} to investigate the adversarial robustness of AVASD and verify the effectiveness of our proposed method to alleviate adversarial attacks.
We reproduce and modify the TalkNet based on the official TalkNet GitHub repository to attack and defend.
To conduct gradient-based adversarial attacks, including BIM, MIM, and PGD, we revise the feature extraction and data augmentation steps using the PyTorch library to make the entire model pipeline differentiable.
For the dataset, we use the AVA Active Speaker dataset \cite{49517}, which contains 29,723 video samples for training.
Since the ground-truth labels for the testing set are not available to the public, we use the validation set with 8,015 samples as the evaluation set.
The lengths of videos range from 1 to 10 seconds and their facial tracks are also provided.
We follow the official evaluation plan and evaluate performance with mean average precision (mAP) \cite{49517}.
The revised TalkNet achieves 92.58\% mAP on the evaluation set, which is slightly higher than the original paper \cite{tao2021someone}.
As the adversarial attack is time- and resource-consuming, we randomly selected 450 genuine samples (225 speaking and 225 non-speaking) with the correct predictions to conduct adversarial attacks. The imperceptible attack budget of audio and visual modality has a very large numerical gap, we introduced $\epsilon_{av}$ to represent the attack budget for easier explanation. The relationships between $\epsilon_{av}$ and attack budget of two modalities are $\epsilon_{a} = \epsilon_{av} \times 10^{-4}$ and $\epsilon_{v} = \epsilon_{av} \times 10^{-1}$, respectively.    

\subsection{The Model Vulnerability under Multi-modality Attacks}
Fig. \ref{fig:multi-attack-ablation} illustrates the attack performance on AVASD under both single-modality and audio-visual attacks, with three attack algorithms in both white-box and black-box attack scenarios. The blue line is the baseline, where the genuine samples are fed directly into the TalkNet model without any attacks.
To investigate the vulnerability of AVASD under both black-box and white-box scenarios, we also trained two models, ncTalNet and specTalkNet. ncTalkNet represents the TalkNet model without the cross attention module, as shown in Fig. \ref{fig:system_overview} (a). specTalkNet denotes the TalkNet by replacing the audio features with linear spectrograms rather than MFCCs adopted by the original TalkNet. 

\noindent
\textbf{White-box and Black-box Attackers.} In Fig. \ref{fig:multi-attack-ablation} (a), there are three settings. 
The TalkNet-TalkNet denotes the white-box scenario, that is, both the model for generating adversarial samples and the target model are TalkNet.
The ncTalkNet-TalkNet and specTalkNet-TalkNet are the black-box scenarios, in which the substitute models for generating adversarial samples are ncTalkNet specTalkNet, respectively, and the target model is TalkNet.  
White-box attackers achieve effective attack performance by degrading the mAP of the TalkNet to a large scale, while black-box attackers can barely manipulate the TalkNet. TalkNet-TalkNet degrades the mAP from 100\% to 65.2\%, when $\epsilon_{av}=5$ under multi-modal attack with PGD method.
But in the same situation, black-box attackers are almost ineffective.

\noindent
\textbf{Single-modal and Multi-modal Attacks.} We take TalkNet-TalkNet to further evaluate the attack performance under single-modal, and multi-modal with the PGD method in Fig. \ref{fig:multi-attack-ablation} (b). When $\epsilon_{av}=5$, the audio-only, visual-only attack, and multi-modal attacks can degrade the model's mAP to 99\%, 77\% and 65.2\%. 
We can observe the same phenomenon in other settings.
As a result, we can derive the following three results:
(1) Audio-only attacks can barely influence the mAP of AVASD. 
(2) Visual-only attacks achieve more effective attack performance than audio-only attacks. One possible reason is that for one video data sample, there are far fewer audio samples than the pixels \cite{szegedy2013intriguing}.
(3) Multi-modal attacks are always more dangerous than single-modal attacks.

\noindent
\textbf{Different Attack Algorithms with Different Attack Budgets.} We take TalkNet-TalkNet to further evaluate the multi-modal attack performance under BIM, MIM, and PGD attacks.
From Fig. \ref{fig:multi-attack-ablation} (c), we have the following observations:
(1) As the attack budget increases, the mAP of TalkNet has a significant decrease trend.
(2) All three attack methods can effectively degrade the AVASD model.
In the following experiments to evaluate the defense performance, we only consider multi-modal attacks in the white-box scenarios, since it is the most dangerous one, and attack budgets are set as $\epsilon_{av}=5$.


\noindent
\textbf{The Imperceptibility of Multi-Modal Attacks.}
We conduct the XAB test to verify that the adversarial noise generated by multi-modal attacks is both visually and acoustically imperceptible.
The XAB test is a standard test to evaluate the detectability between two sensory stimulus choices. 
We randomly select 4 adversarial-genuine pairs (2 speech and 2 non-speech pairs) for each of the three attack algorithms with $\epsilon_{av}=5$, resulting in 12 pairs of randomly selected adversarial-genuine pairs (i.e., A and B).
One reference data (i.e., X) is chosen from A and B. 
A, B, and X are shown to the volunteers.
The volunteers should select the more similar data to X, from A and B.
We hire five volunteers to join in the XAB test. 
The classification accuracy for the XAB test is 53.33\%, which is a nearly random guess, leading to the conclusion that the adversarial samples are difficult to be distinguished from genuine samples. 
The XAB test samples will be shown here \footnote{https://xjchengit.github.io/Push-Pull/index.html}.

\begin{table*}[ht]
    \setlength{\tabcolsep}{3.0pt}
    \renewcommand\arraystretch{1.0}
    \fontsize{8}{10}\selectfont
    \center
    \begin{tabular}{lc|c|c|ccc|ccc|ccc} 
\hline
     &\multirow{3}{*}{\shortstack[c]{Model}} & \multirow{3}{*}{\shortstack[c]{Adversarial \\ training \cite{goodfellow2014explaining}}} & \multirow{3}{*}{\shortstack[c]{Clean \\ mAP (\%)}} &    \multicolumn{9}{c}{ Different attack methods with $\mathcal{L}_{{CE}_{all}}$}    \\
\cline{5-13}
&&&& \multicolumn{3}{c|}{BIM} & \multicolumn{3}{c|}{MIM} & \multicolumn{3}{c}{PGD} \\

\cline{5-13}
&&&& A (ECR) & V (ECR) & mAP (\%) & A (ECR) & V (ECR) & mAP (\%) & A (ECR) & V (ECR) & mAP (\%) \\

\hline
(A) & $\mathcal{L}_{{CE}_{all}}$  \ 
                                & \XSolidBrush & 92.58   &  0.1658   &  0.3140   &  49.53  \ 
                                                         &  0.1683   &  0.3170   &  49.30  \ 
                                                         &  0.1715   &  0.3236   &  47.79  \\
\hline
(B1) & $\mathcal{L}_{{CE}_{all}}$ \ 
                                    & BIM   & 92.15  &  0.2759  &  0.2644  &  62.7   \ 
                                                 &  0.2778  &  0.2663  &  59.26   \ 
                                                 &  0.2937  &  0.2772  &  60.01   \\
(B2) & $\mathcal{L}_{{CE}_{all}}$ \ 
                                     & MIM &  91.34 &  0.3052  &  0.2108  &  54.66   \ 
                                                 &  0.3073  &  0.2133  &  52.18  \ 
                                                 &  0.3030  &  0.2118  &  54.23 \\
(B3) &  $\mathcal{L}_{{CE}_{all}}$  \ 
                                & PGD  & 91.68  &  0.2728  &   0.1846    &    58.29  \ 
                                                &  0.2783  &   0.1893    &    58.3  \ 
                                                &  0.2840  &   0.1938    &    56.06  \\
\hline
(C1) & $\mathcal{L}_{{CE}_{all}} + \mathcal{L}_{1}$ \ 
                                & \XSolidBrush & 92.09  &  0.1407  &  0.2603   &  82.96  \ 
                                                        &  0.1382  &  0.2618   &  81.32  \ 
                                                        &  0.1379  &  0.2677   &  80.98  \\
(C2) & $\mathcal{L}_{{CE}_{all}} + \mathcal{L}_{2}$ \ 
                              & \XSolidBrush   &  92.05 &   0.1451   &  0.1444  &  92.65  \ 
                                                        &  0.1496    &  0.1481  &  90.69  \ 
                                                        &  0.1501    &  0.1509  &  88.93  \\
(C3) & $\mathcal{L}_{{CE}_{all}} + \mathcal{L}_{3}$\ 
                              & \XSolidBrush & 92.16  &  0.1575  &  0.3264  &  74.98 \ 
                                                     &  0.1602   & 0.3289   &  76.25 \ 
                                                     &  0.1590   & 0.3343   &  73.97 \\
(C4) & $\mathcal{L}_{{CE}_{all}} + \mathcal{L}_{4}$ \ 
                          & \XSolidBrush &  91.28 & 0.1065  &  0.2262   & 83.82 \ 
                                                  & 0.1154  &  0.2322   & 79.82 \ 
                                                  & 0.1158  &  0.2351   & 78.72 \\
\hline
(D1) & $\mathcal{L}_{{CE}_{all}} + \mathcal{L}_{1}  + \mathcal{L}_{2}$ \ 
                            & \XSolidBrush & 92.46  &  0.2182   &  0.4672  &  66.91  \ 
                                                    &  0.2149   &  0.4656  &  67.89  \ 
                                                    &  0.2317   &  0.4910  &  64.11 \\
(D2) & $\mathcal{L}_{{CE}_{all}} + \mathcal{L}_{1}  + \mathcal{L}_{3}$ \ 
                      & \XSolidBrush   &  92.20  &  0.2218   &  0.4102   &  48.16  \ 
                                                 &  0.2190   &  0.4134   &  47.92  \ 
                                                 &  0.2239   &  0.4228   &  49.27  \\
(D3) & $\mathcal{L}_{{CE}_{all}} + \mathcal{L}_{1}  + \mathcal{L}_{4}$ \ 
                            & \XSolidBrush  &  91.81  &  0.0820   &  0.2194  &  93.86  \ 
                                                  &  0.0834   &  0.2337   &   93.34    \ 
                                                  &  0.0811   &  0.2313   &   93.15    \\
(D4) & $\mathcal{L}_{{CE}_{all}} + \mathcal{L}_{2}  + \mathcal{L}_{3}$ \ 
                            & \XSolidBrush  &  92.27 &  0.1525   &  0.3094   &   57.02  \ 
                                                     &  0.1500   &  0.3029   &   63.36  \
                                                     &  0.1549   &  0.3135   &   61.54  \\
(D5) & $\mathcal{L}_{{CE}_{all}} + \mathcal{L}_{2}  + \mathcal{L}_{4}$ \ 
                            & \XSolidBrush  &  91.93 &  0.0936   &  0.1583  &  68.12  \ 
                                                     &  0.0962   &  0.1612  &  66.28  \ 
                                                     &  0.0992   &  0.1667  &  67.75     \\
(D6) & $\mathcal{L}_{{CE}_{all}} + \mathcal{L}_{3}  + \mathcal{L}_{4}$ \
                            & \XSolidBrush &  91.70 &   0.0782  &  0.2128   &  91.79     \ 
                                                  &  0.0771   & 0.2135  &  92.48 \ 
                                                  &  0.0785   & 0.2172  &  91.01      \\
\hline

\hline
(E1) &  $\mathcal{L}_{{CE}_{all}} + \mathcal{L}_{1}  + \mathcal{L}_{4}$ \ 
                            &   BIM    & 90.63  &  0.0989  &  0.1007  &  97.85   \ 
                                                &  0.1006  &  0.1011  &  97.6   \ 
                                                &  0.0955  &  0.1040  &  97.47   \\
(E2) &  $\mathcal{L}_{{CE}_{all}} + \mathcal{L}_{1}  + \mathcal{L}_{4}$ \ 
             &   MIM   &  91.70 &  \textbf{0.0344}  &  \textbf{0.0676}  &  \textbf{99.99}   \ 
                                                &  \textbf{0.0341}  &  \textbf{0.0669}  &  \textbf{99.98}  \ 
                                                &  \textbf{0.0355}  &  \textbf{0.0696}  &  \textbf{99.97} \\
(E3) &  $\mathcal{L}_{{CE}_{all}} + \mathcal{L}_{1}  + \mathcal{L}_{4}$ \ 
                             &  PGD   & 91.88  &  0.0470  &  0.1001  &  97.68  \ 
                                              &  0.0446  &  0.0966  &  97.47  \ 
                                              &  0.0423  &  0.0953  &  98.67  \\
\hline

\vspace{-1.5em}
\end{tabular}
\caption{AVASD performance of different models under three attack algorithms.}
\label{tab:ablation_study}
\vspace{-1em}
\end{table*}

\subsection{Audio-Visual Interaction Loss}
Table~\ref{tab:ablation_study} shows the defense performance of different methods under three attack algorithms.
(A) denotes the original TalkNet model.
(B1)-(B3) are the baselines, which represent models trained using adversarial training \cite{goodfellow2014explaining}, and the adversarial examples are generated by BIM, MIM, and PGD, respectively. Adversarial training is conducted by injecting the adversarial data into the training set and thus alleviating the adversarial invulnerability.
(C1)-(C4) denote the model trained by incorporating only one of the four losses in Section~\ref{sec:avil}.
We exhaust the pairwise permutation of the four losses to train AVASD models, which are shown in (D1)-(D6).
We select the model with the best defense performance from (D1)-(D6), namely (D3), and combine it with adversarial training to see whether the AVIL can complement adversarial training, and the models are shown in (E1)-(E3).
The clean mAP(\%) column is the mAP performance testing on the entire evaluation set without adversarial attacks.
In order to conduct fair comparison, we get the data with correct prediction for model (A)-(E3), and do intersection of such data to get the testing data.

To show the effectiveness of defense methods against adversarial noise, we also set up another evaluation metric, the embedding change ratio (ECR) for audio embedding $z_{a}$ and visual embedding $z_{v}$ after the cross-attention as shown in Fig.~\ref{fig:system_overview}. (a).
Take the audio ECR for example, it is calculated by:
$1/K \times \sum^{K}_{i=1} ||z_{a,i} - \tilde{z}_{a,i}||_2/|| z_{a,i} ||_2,$
where $\tilde{z}_{a,i}$ and $z_{a,i}$ are the adversarial and genuine embeddings, $K$ is the number of total frames. 
ECR measures the embedding change ratio before and after adversarial attacks. 
A (ECR) and V (ECR) denote the ECR of the audio and visual parts, respectively.
The lower the ECR is, the less effect introduced by the attack algorithms and the better the defense performance will be.

\noindent
\textbf{Baselines.}
From (A), the original TalkNet model performs well on the AVASD task with 92.58\% mAP for the clean samples, as shown in Table~\ref{tab:ablation_study}.
However, the multi-modality attacks seriously degrade the mAP of (A).
It can only get 49.53\%, 49.40\%, and 47.79\% mAP under BIM, MIM, and PGD attack algorithms respectively.
According to the (B1)-(B3) of Table~\ref{tab:ablation_study}, adversarial training does improve the robustness of the AVASD model.
Using the BIM attack algorithm to generate adversarial examples and then conducting adversarial training achieves the best defense performance compared with the other two attack algorithms, resulting in 13.17\%, 9.96\%, and 12.22\% absolute improvement of mAP under BIM, MIM, and PGD attacks, respectively.
In terms of ECR, adversarial training effectively reduce V(ECR), yet significantly increase A(ECR).

\noindent
\textbf{Using One AVIL.}
As shown in (C1)-(C4), using any one of the four AVILs improves the mAP to a large scale, resulting in better defense performance the (B1)-(B3), the adversarial training using BIM, MIM, and PGD.
(C2), using $\mathcal{L}_{2}$ leads to the best performance.
It seems that when only introducing one loss, maximizing the intra-modality and intra-class similarity is the best choice to tackle the adversarial noise.
Regarding the ECR, (C1)-(C4) help reduce A(ECR) and V(ECR) in most of the settings.

\noindent
\textbf{Pairwise Permutations of AVILs.}
From (C1)-(D6) in Table~\ref{tab:ablation_study}, we have the following observations and analysis: 
(1) Adopting pairwise permutations of AVILs perform better than adversarial training in most of the settings from both mAP and ECR perspectives. 
(2) Combining $\mathcal{L}_{1}$ and $\mathcal{L}_{4}$ even improves the mAP to 93.86\%, 93.34\%, and 93.15\% for BIM, MIM, and PGD respectively. It is the most robust combination.
A possible reason is that enlarging inter-class dispersion (minimizing $\mathcal{L}_{1}$) and maximizing intra-class similarity (maximizing $\mathcal{L}_{4}$) at the same time will result in the model with the best robustness.

\noindent
\textbf{Integrating AVIL with Adversarial Training.}
(E1)-(E3) show that combining AVIL with adversarial training can leverage their complementary to improve the adversarial robustness.
Furthermore, integrating AVIL with MIM-based adversarial training can improve the mAP to over 99\% under BIM, MIM, and PGD attacks. 

\begin{figure}
\centering
\includegraphics[width=8.cm]{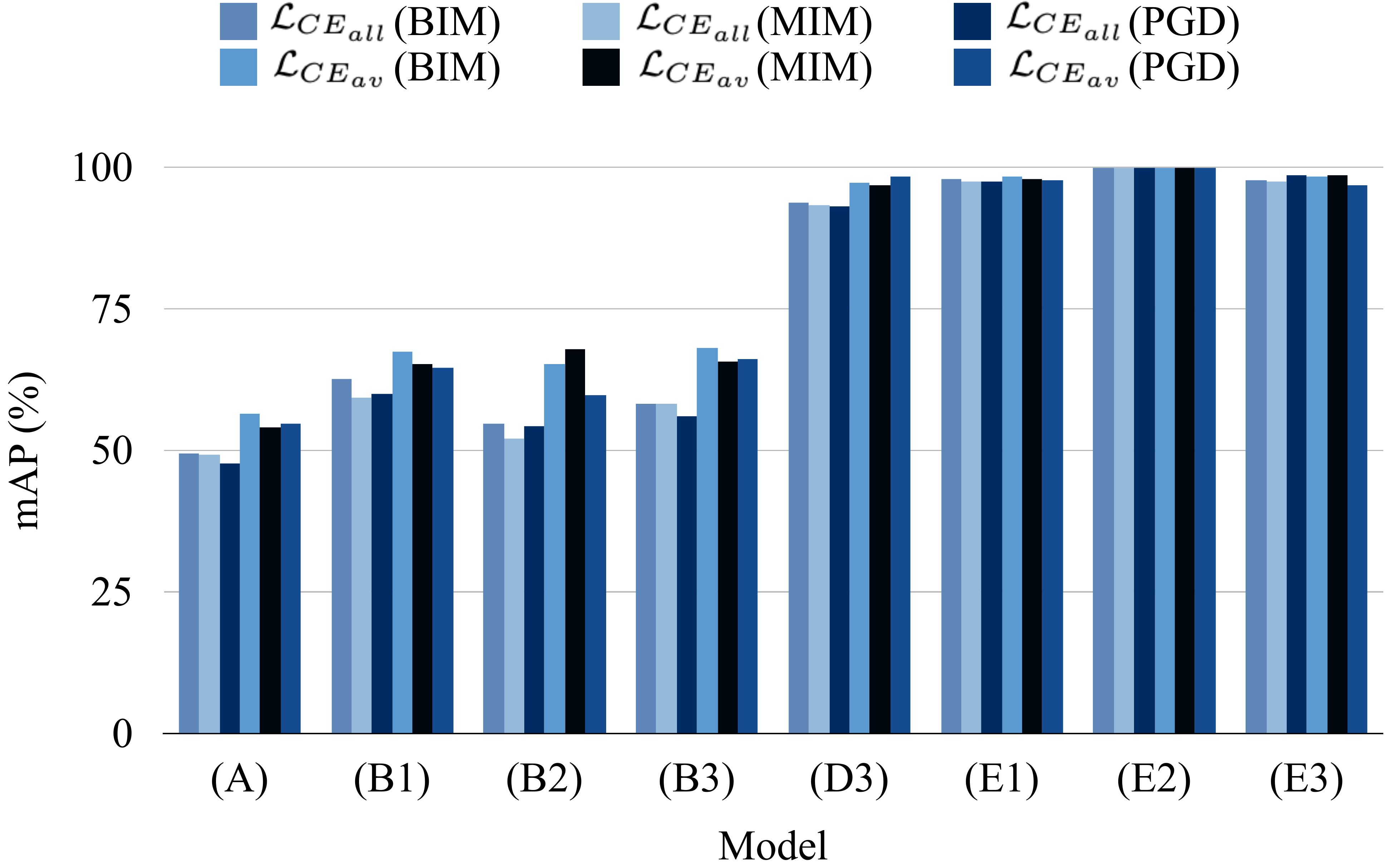}
\vspace{-1.3em}
\caption{Training-aware ($\mathcal{L}_{{CE}_{all}}$) attack and inference-aware ($\mathcal{L}_{{CE}_{av}}$) attack scenarios.}
\label{fig:different_scenario}
\end{figure}

\subsection{Training-aware and Inference-aware Attacks}
\label{sec:infer vs train}
Fig.~\ref{fig:different_scenario}. compares the performance of 8 AVASD models in Table \ref{tab:ablation_study} under the training-aware and inference-aware scenarios with BIM, MIM, PGD attack method. We also have the same evaluation set as Table \ref{tab:ablation_study}. The legend show the two scenarios with three attack method. For instance, the legend ``$\mathcal{L}_{CE_{all}}$ (BIM)" denotes using the $\mathcal{L}_{CE_{all}}$ to craft the adversarial samples with BIM attack method. The $L_{CE_{all}}$ and $L_{CE_{av}}$ are defined in equations \ref{eq:CE all} and \ref{eq:CE av}.
We have the following observations and analysis:
(1) In (A) groups, we can see that the performance of the AVASD model has also dropped significantly in the inference-aware scenario, but the inference-aware attacks are less dangerous compared with training-aware attacks.
(2) From (B1)-(B3), adversarial training does alleviate the adversarial noise with the same trend in the training-aware attack scenario.
(3) Compare (D3) with (B1)-(B3), the AVIL performs better in improving the adversarial robustness than adversarial training.
(4) From (E1)-(E3), we can see that AVIL can complement adversarial training.
To sum up, we can conclude that the inference-aware attack scenario has the same trend as the training-aware attack scenario.

\section{Conclusion}
In this work, we first expose that audio-visual active speaker detection models are highly susceptible to adversarial attacks through comprehensive experiments, by investigating the white-box and black-box adversaries, single- and multi-modal attacks, training-aware, and inference-aware attack scenarios, and three attack algorithms with several attack budgets.
Also, we propose the audio-visual interaction loss to enlarge the inter-class difference and intra-class similarity, resulting in more robust AVASD models for which budget-limited attackers can not find feasible adversarial samples.
The experimental results illustrate that the proposed method is far more effective than adversarial training, and the proposed AVIL can complement adversarial training to further alleviate the adversarial vulnerability of AVASD models.
In the future, we will investigate hyperparameter searching strategies to further improve the effectiveness of the proposed AVIL.


\section{ACKNOWLEDGMENTS}
\label{sec:ack}
This work was partially supported by the Ministry of Science and Technology, Taiwan (Grant no. MOST109-2221- E-002-163-MY3), and the Centre for Perceptual and Interactive Intelligence, an InnoCentre of The Chinese University of Hong Kong. This work was done while Haibin Wu was a visiting student at CUHK. Haibin Wu is supported by Google PHD Fellowship Scholarship.

\bibliographystyle{IEEEbib}
\bibliography{strings,refs}

\end{document}